\documentclass[longauth]{aa} 

\usepackage{graphicx}
\usepackage{txfonts}

\newcommand{\kms}{km\,s$^{-1}$}

\usepackage{hyperref}
\hypersetup{colorlinks=true,allcolors=[rgb]{0,0,0.8}}

\begin{document}

\title{Disk Evolution Study Through Imaging of Nearby Young Stars (DESTINYS): Evidence of planet-disk interaction in the 2MASSJ16120668-3010270 system\thanks{Based on observations made with ESO telescopes at the La Silla Paranal Observatory under program IDs 111.255B.001, 1104.C-0415(D) and 109.23BC.001.}}
\titlerunning{Planet-disk interaction in 2MASSJ1612}


\author{C. Ginski\inst{1}
          \and
        P. Pinilla\inst{2}
        \and
        M. Benisty\inst{3,4}
        \and 
        C. Pinte\inst{5,4}
        \and
        R. Claes\inst{6}
        \and
        E. Mamajek\inst{7}
        \and
        M. Kenworthy\inst{8}
        \and
        M. Murphy\inst{1}
        \and
        C. Manara\inst{6}
        \and 
        J. Bae\inst{9}
        \and
        T. Birnstiel\inst{10,11}
        \and
        J. Byrne\inst{1}
        \and
        C. Dominik\inst{12}
        \and
        S. Facchini\inst{13}
        \and
        A. Garufi\inst{14}
        \and
        R. Gratton\inst{15}
        \and
        M. Hogerheijde\inst{8,12}
        \and
        R. van Holstein\inst{16}
        \and
        J. Huang\inst{17}
        \and 
        M. Langlois\inst{18}
        \and
        C. Lawlor\inst{1}
        \and
        J. Ma\inst{4}
        \and
        D. McLachlan\inst{1}
        \and
        F. Menard\inst{4}
        \and
        R. Rigliaco\inst{15}
        \and
        A. Ribas\inst{19}
        \and 
        T. Schmidt\inst{20}
        \and
        A. Sierra\inst{2}
        \and
        R. Tazaki\inst{4}
        \and
        J. Williams\inst{21}
        \and
        A. Zurlo\inst{22,23}
          }

\institute{School of Natural Sciences, Center for Astronomy, University of Galway, Galway, H91 CF50, Ireland\\
              \email{christian.ginski@universityofgalway.ie}
              \and
                Mullard Space Science Laboratory, University College London, Holmbury St Mary, Dorking, Surrey RH5 6NT, UK
              \and Universit\'{e} C\^{o}te d'Azur, Observatoire de la C\^{o}te d'Azur, CNRS, Laboratoire Lagrange, Bd de l'Observatoire, CS 34229, F-06304 Nice cedex 4, France
             \and Universit\'{e} Grenoble Alpes, CNRS, Institut de Plan\'{e}tologie et d’Astrophysique (IPAG), F-38000
Grenoble, France
            \and
            School of Physics and Astronomy, Monash University, Clayton Vic 3800, Australia
            \and
            European Southern Observatory, Karl-Schwarzschild-Strasse 2,
85748, Garching bei München, Germany
\and
Jet Propulsion Laboratory, California Institute of Technology, 4800 Oak Grove Drive, Pasadena, CA 91109, USA
\and
Leiden Observatory, Leiden University, P.O. Box 9513, 2300 RA Leiden, The Netherlands
\and
Department of Astronomy, University of Florida, Gainesville, FL, 32611, USA
\and
University Observatory, Faculty of Physics, Ludwig-Maximilians-Universität München, Scheinerstr. 1, 81679 Munich, Germany
        \and
        Exzellenzcluster ORIGINS, Boltzmannstr. 2, D-85748 Garching, Germany
\and
Anton Pannekoek Institute for Astronomy, University of Amsterdam, Science Park 904, 1098 XH Amsterdam, The Netherlands
\and
Dipartimento di Fisica, Universit\`a degli Studi di Milano, Via Celoria, 16, Milano, I-20133, Italy
\and
INAF, Osservatorio Astrofisico di Arcetri, Largo Enrico Fermi 5, I-50125, Firenze, Italy
\and
INAF-Osservatorio Astronomico di Padova, Vicolo dell'Osservatorio 5, 35122, Padova, Italy
\and
European Southern Observatory, Alonso de C\'{o}rdova 3107, Vitacura, Casilla 19001, Santiago, Chile  
\and Department of Astronomy, Columbia University, 538 W. 120th Street, Pupin Hall, New York, NY, United States of America
\and
Centre de Recherche Astrophysique de Lyon, CNRS, UCBL, ENS Lyon, UMR 5574, F-69230, Saint-Genis-Laval, France 
\and
Institute of Astronomy, University of Cambridge, Madingley Road,
Cambridge CB3 0HA, UK
\and
Hamburger Sternwarte, Gojenbergsweg 112, 21029, Hamburg, Germany
\and
Institute for Astronomy, University of Hawai’i at Manoa, Honolulu, HI 96822, USA
\and
Instituto de Estudios Astrof\'isicos, Facultad de Ingenier\'ia y Ciencias, Universidad Diego Portales, Av. Ej\'ercito Libertador 441, Santiago, Chile
              \and
              Millennium Nucleus on Young Exoplanets and their Moons (YEMS), Chile 
             }

\date{Received September 15, 1996; accepted March 16, 1997}

 
\abstract
   {The multitude of different architectures found for evolved exoplanet systems are in all likelihood set during the initial planet-formation phase in the circumstellar disk. To understand this process, we have to study the earliest phases of planet formation.}
   {Complex sub-structures, believed to be driven by embedded planets, have been detected in a significant portion of the disks observed at high angular resolution. We aim to extend the sample of such disks to low stellar masses and to connect the disk morphology to the expected proto-planet properties.}
   {In this study, we used VLT/SPHERE to obtain resolved images on the scale of $\sim$10\,au of the circumstellar disk in the 2MASSJ16120668-3010270 system in polarized scattered light. We searched for the thermal radiation of recently formed gas giants embedded in the disk. Additionally, we used VLT/XSHOOTER to obtain the stellar properties in the system.}
   {We resolve the disk in the 2MASSJ16120668-3010270 system for the first time in scattered near-infrared light and reveal an exceptionally structured disk. We find an inner disk (reaching out to 40\,au) with two spiral arms, separated by a gap from an outer ring extending to 115\,au. By comparison with our own model and hydrodynamic models from the literature, we find that these structures are consistent with the presence of an embedded gas giant with a mass range between 0.1\,M$_\mathrm{Jup}$ and 5\,M$_\mathrm{Jup}$ depending on the employed model and their underlying assumptions. Our SPHERE observations find a tentative candidate point source within the disk gap, the brightness of which would be consistent with this mass range if it indeed traces thermal emission by an embedded planet. This interpretation is somewhat strengthened by the proximity of this signal to compact millimeter continuum emission in the disk gap, which may trace circumplanetary material. It is, however, unclear if this tentative companion candidate could be responsible for the observed disk gap size, given its close proximity to the inner disk.  Generally, our VLT/SPHERE observations set an upper limit of $\sim$5\,M$_\mathrm{Jup}$ in the disk gap ($\sim$0.2"-0.5"), consistently with our modeling results. 
   
   The 2MASSJ16120668-3010270 system is one of only a few systems that shows this exceptional morphology of spiral arms located inside a scattered light gap and ring. We speculate that this may have to do with a higher disk viscosity compared with other systems such as PDS\,70. If planets in the disk are confirmed, 2MASSJ16120668-3010270 will become a prime laboratory for the study of planet-disk interaction.}
   {}


   \maketitle
%

\section{Introduction}
High-angular-resolution images obtained with ALMA \citep[e.g.,][]{Andrews2020} and in near-infrared scattered light \citep[e.g.,][]{Benisty2023} have consistently shown a significant occurrence rate of disk substructures, appearing as cavities, rings, gaps, asymmetries, and spiral arms. 
The analysis of the substructures' properties can directly inform us of the local disk's physical conditions. For example, spiral arms can be used to derive the disk's thermal structure \citep{Rosotti2020}, while rings can be used to understand the disk's vertical structures \citep{Avenhaus2018}. Multi-wavelength observations of substructures enable constraints on the dust-grain size distribution \citep{Sierra2021} and the potential for planetesimal formation \citep{Stammler2019,Zagaria2023}.
However, due to observational requirements the overall observed disk sample is biased toward bright and large disks \citep{Garufi2022,Bae2023}, which makes the inference of general patterns in the early planet-formation process challenging. 

In particular, scattered light observations and survey were often limited to early-type stars (e.g., GEMINI LIGHTS, \citealt{Rich2022}) due to the need for sufficient optical flux to drive modern extreme-adaptive-optics instruments. The Disk Evolution Study Through Imaging of Nearby Young Stars (DESTINYS, \citealt{Ginski2020, Ginski2021}) is an ESO large program that specifically aims to push the current observational limits of instrumentation by observing late-type stars down to $\sim$0.5\,M$_\odot$ in nearby star-forming regions. In recent survey papers for Taurus, Chamaeleon, and Orion (\citealt{Garufi2024, Ginski2024, Valegard2024}), we presented the first systematic studies of each of these star-forming regions in scattered light. One of the key findings is that older transition disks are generally brighter in scattered light and show an intricate substructure. These results show that transition disks (disks with dust-depleted inner cavities) are ideal targets for the study of substructures.
Although pinpointing the exact origin of substructures remains challenging - with the exception of a few cases for which a stellar companion \citep{Gonzalez2020} or giant planets \citep{Keppler2018} are observed - the morphology of the substructures can in principle also provide a hint as of what is causing them: spiral arms seem to naturally arise in the interactions between the disk and a (planetary or stellar) companion \citep[e.g.,][]{Baruteau2014}, while multiple rings and gaps (the most common type of substructures) can have both planetary and non-planetary origins \citep{Bae2023, Lesur2023}. 

Motivated by the discovery of PDS\,70b and c, transition disks have been surveyed with direct imaging campaigns to search for embedded planets \citep{Ren2023}. The typical mass-detection limit for these objects lies within a few Jupiter masses outside of the disk's signal \citep{asensio2021}. While the unambiguous confirmation of new candidate planets remains challenging, this subgroup of disks shows complex substructures, with a clear prevalence of spiral-arm-hosting disks around intermediate-mass stars with strong near-infrared excess \citep{Garufi2018, Benisty2023}, which is indicative of a highly perturbed inner-disk component. 


In this paper, we report new scattered-light observations of a PDS\,70-like transition disk around a M0.5 young star, 2MASSJ16120668-3010270 (also known as RIK 113 in some literature; hereafter referred to as 2MASSJ1612), which is located in the Sco-Cen association at a distance of 132.1 pc \citep{Gaia_2023}. This is one of the lowest mass central stars observed with the remit of DESTINYS. Recently, this disk was resolved for the first time with submillimeter ALMA observations (\citealt{Sierra2024}), which show a prominent disk cavity and bright outer ring. \cite{Sierra2024} also reported the detection of compact emission, possibly tracing circumplanetary material, in the disk cavity. This makes 2MASS1612 a prime candidate for high-resolution follow-up observations.
The paper is structured as follows. Section\,\ref{sec:prev_work} summarizes previous studies of the 2MASSJ1612 system, Section\,\ref{sec:obs} describes the VLT/SPHERE and VLT/XSHOOTER observations, Section\,\ref{sec:stellar} the derived stellar properties, and Section\,\ref{sec:disk} the analysis of the disk image. Section\,\ref{section:discussion} features a discussion of the results. Section\,\ref{sec:summary} provides our conclusions. 

\section{Previous studies of the 2MASSJ1612 system} \label{sec:prev_work}

The target star was first detected as an H$\alpha$ emission star in the objective prism survey of \citet{MacConnell1981} (IDed in SIMBAD as [M81] I-497). 
The star remained anonymous until a kinematic and photometric survey was carried out by \citet{Rizzuto2015} using the UCAC4, APASS, 2MASS, and WISE surveys \citep{Zacharias2013,Henden2012,Cutri2003,Cutri2012}, thereby selecting the star 2MASS J16120668-3010270 as a candidate member of the Upper Sco association \citep[e.g.,][]{Preibisch2008,Pecaut2012}. 
Spectroscopic follow-up by \citet{Rizzuto2015} with the WiFeS instrument on the ANU 2.3 m found the star to be a lightly reddened ($A_V$ = 0.2 mag) M0.5-type star with strong H$\alpha$ emission (EW(H$\alpha$) $\simeq$ -34 \AA) and strong Li absorption (EW(Li$\lambda$6708) $\simeq$ 0.55 \AA).
\citet{Rizzuto2015} found the WISE W2/W3/W4 photometry (4.6$\mu$m, 11.6$\mu$m, 22.1$\mu$m) to all have excess emission consistent with a "full disk".
Hence, all the indicators (kinematic, photometric, spectroscopic) were consistent with the star being a pre-main-sequence "Classical T Tauri" member of Upper Sco, and the star was designated "RIK 113".\\

After its inclusion as a likely member of Upper Sco, RIK 113 appeared in several more surveys.
\citet{Kuruwita2018} analyzed five spectra for RIK 113 using WiFeS, noting its variation in H$\alpha$ EW (ranging between -19.6\,$\AA$ and -30.2\,$\AA$) and reporting five radial velocities whose distribution was statistically consistent with the star being single. 
A high-resolution spectroscopy survey of Upper Sco members by \citet{Fang2023} reported [O I]$\lambda$ 6300 emission (FWHM = 23.8 \kms, EW([O I]$\lambda$6300) = 0.28\,$\AA$) characterized by a single Gaussian component, which is consistent with a disk wind.
A speckle imaging survey with SOAR by \citet{Tokovinin2020} found no evidence of a stellar companion. 
With the advent of the Gaia data releases \citep{GaiaDR1,GaiaDR2,GaiaDR3}, analyses of RIK 113's astrometry led multiple teams to assign it to Upper Sco using Gaia DR1 \citep{Galli2018,Luhman2018}, DR2 \citep{Damiani2019,Luhman2020}, and DR3 \citep{Kerr2021}.
\cite{Damiani2019} assigned RIK 113 to a kinematic subpopulation of Upper Sco dubbed D2a, whereas \citet{Luhman2020} and \citet{Kerr2021} assigned it to the general Upper Sco population.
\citet{Luhman2020} estimated the mean age of Upper Sco to be 10.5 Myr (non-magnetic evolutionary tracks) or 12 Myr (magnetic evolutionary tracks), and \citet{Kerr2021} estimated a mean group age of 11.3 Myr.
These Gaia-informed estimates are consistent with the previous isochronal age estimate based on the $\sim$1-17\,$M_{\odot}$ membership from \citet{Pecaut2012} of $11(\pm1\pm2)$ Myr (statistical, systematic) and the updated estimate of $10\pm3$ Myr from \citet{Pecaut2016}, which found an age gradient across US, with a $\sigma$ age spread of $\pm7$ Myr across the group.

\section{Observations}
\label{sec:obs}

We obtained data from the ESO Very Large Telescope, namely SPHERE (\citealt{Beuzit2019}) imaging observations of 2MASSJ1612, as well as XSHOOTER \citep{vernet2011} spectroscopic observations. In the following, we briefly summarize the instrument setup and subsequent data processing. 

\subsection{SPHERE imaging}

The star 2MASSJ1612 was observed with the IRDIS (\citealt{Dohlen2008}) subsystem of SPHERE on April 12, 2023 and on June 21, 2023. Observations at both epochs were performed in the polarimetric imaging mode of IRDIS  (\citealt{deBoer2020, vanHolstein2020})  in two different filters. The earlier epoch was observed in the H band, while the later epoch was observed in the K band. In both cases the instrument was in pupil-tracking mode, allowing the field of view to rotate. The central star was placed behind the SPHERE apodized Lyot coronagraph (radius of 92.5\,mas). The H-band observation sequence in April 2023 recorded a total of 104 images with an individual exposure time of 32\,s. The images were split into 26 polarimetric cycles (with 4 half-wave-plate orientations per cycle). Conditions were excellent, with an average seeing of 0.5'' and an atmosphere coherence time of 9\,ms. The K-band observations of June 2023 were additionally taken in star-hopping mode (\citealt{Wahhaj2021}); i.e., reference-star images were interspersed with the science observations. We recorded a total of 52 science images with an individual exposure time of 64\,s. The science observations were split between 13 polarimetric cycles. Seeing conditions were similar to those for the previous epoch, but coherence time was variable between 4\,ms and 8\,ms. \\

All IRDIS data were reduced with the IRDAP (\citealt{vanHolstein2020}) pipeline using default settings\footnote{\url{https://irdap.readthedocs.io}}. The final products included polarized scattered-light images of the disk after application of polarization differential imaging (\citealt{Kuhn2001}), as well as total-intensity images after application of angular differential imaging (\citealt{Marois2006}). Furthermore, we used the reference-star images of our K-band observation epoch to perform reference differential imaging. To minimize over-subtraction of disk signal, in this case we followed the procedure described in \cite{Ginski2021}, which utilizes an iterative process to subsequently subtract disk signal from the image before the fitting of the stellar signal.

\subsection{XSHOOTER spectroscopy}

We obtained one spectrum with the VLT/X-Shooter instrument on May 18, 2023 (Pr.Id. 111.255B.001, PI Claes). The observations were carried out under clear weather conditions and fully within constraints. 
The X-Shooter spectrum simultaneously covers the wavelength region between $\sim$300 and 2500\,nm, with a resolution of R$\sim$5400, 18400, and 11600 in the UVB ($\lambda\sim300-550$\,nm), VIS ($\lambda\sim500-1050$\,nm), and NIR ($\lambda\sim1000-2500$\,nm) arms, respectively. This resolution was obtained using the slit width of 1.0'', 0.4'', and 0.4'' in the three arms. Prior to the observations with this set of slits, a short exposure with 5.0'' wide slits was taken to allow the narrow slit observations to be corrected for slit losses.

Data reduction was carried out with the X-Shooter pipeline  v.4.2.2 \citep{xspipe} in the Reflex workflow \citep{reflex}. The pipeline carries out the reduction steps including order combination, rectification, flux calibration, and extraction of the 1D spectrum. The latter was then absolutely flux calibrated by rescaling the flux to the one obtained with the wide slit, following commonly used procedures \citep[e.g.,][]{Manara2021}. Telluric correction was performed using the {\tt molecfit} tool \citep{molecfit1}.

\section{Stellar properties}\label{sect::star_prop}
\label{sec:stellar}

We derived the stellar and accretion properties of 2MASSJ1612 by fitting the X-Shooter spectrum using the method described by \citet{manara13} and recently improved by \citet{claes24}. The observed spectrum is fit using a grid of photospheric templates obtained by interpolating spectra of non-accreting young stellar objects, a reddening law, and a slab model to reproduce the accretion shock. 

\begin{figure}
\centering
    \includegraphics[width=0.5\textwidth]{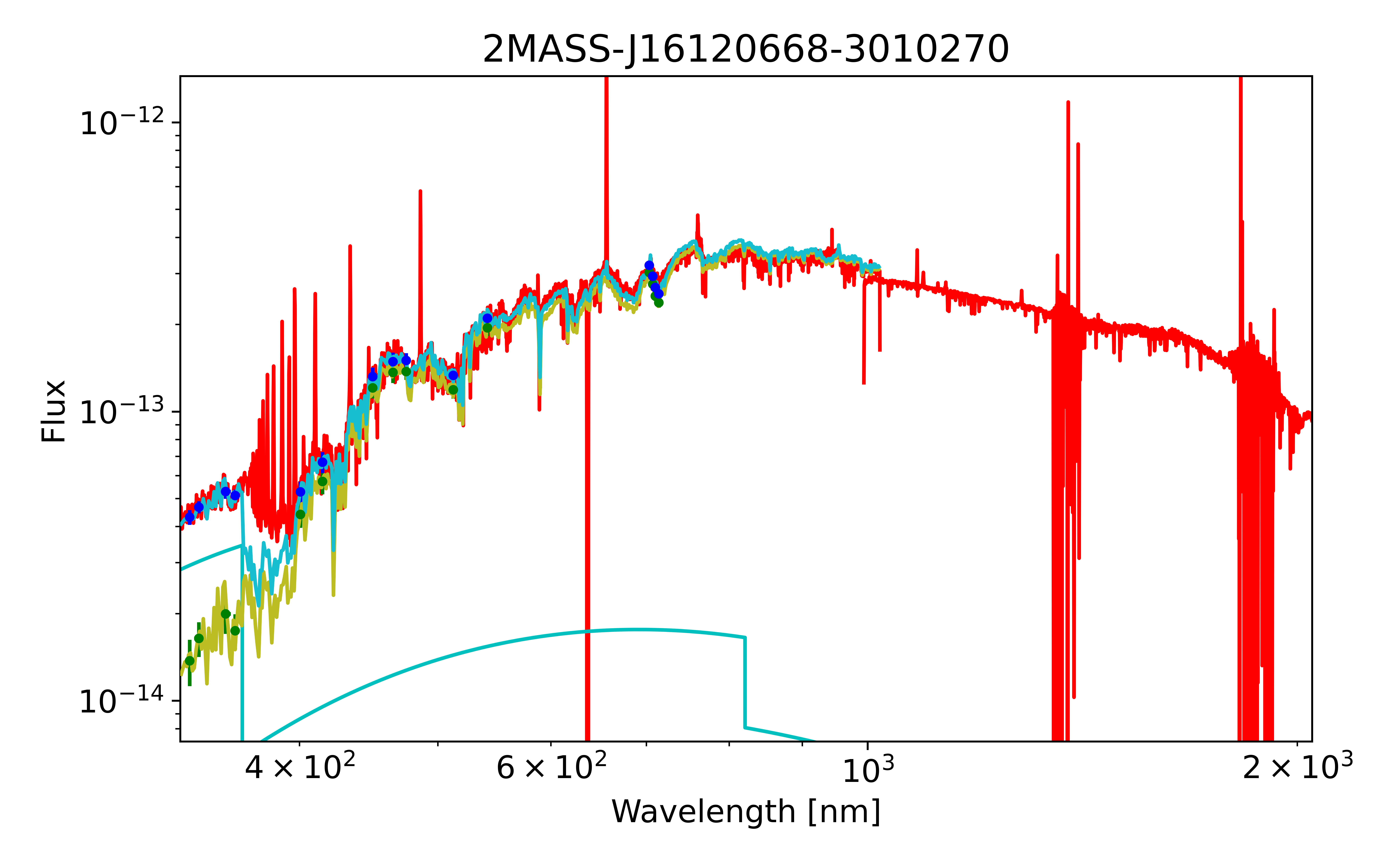}
    \caption{Best fit of the X-Shooter spectrum of  2MASSJ1612. The observed spectrum is shown in red, the best-fit photospheric template in yellow, and the slab model in cyan. The best fit is shown in light blue and reproduces the observed spectrum from $\sim$320 nm to $\sim1\mu$m. }
    \label{fig::best_fit_XS}
\end{figure}

The best fit for our target is shown in Figure~\ref{fig::best_fit_XS} and is obtained using a photospheric template with spectral type M0.5, $A_V$ = 0 mag, and a slab model corresponding to $\log(L_{\rm acc}/L_\odot)$ = -2.092. This fit results in an estimate of $M_\star = 0.60\pm0.05 M_\odot$ (very similar to the stellar mass of 0.7\,M$_\odot$ from the ALMA $^{12}$CO velocity-map fits; \citealt{Sierra2024}) and $\dot{M}_{\rm acc} = 6.1\times10^{-10} M_\odot$/yr using the evolutionary models by \citet{Baraffe2015}. We note that another solution with $A_V$ = 0.4 mag did not fit the spectrum in the reddest part of the VIS arm, and the solution with no extinction is preferred. The inferred age from isochrone interpolation is $\sim$5 Myr. The other inferred parameters are reported in Table~\ref{tab::best_fit_XS}.

\begin{table}[t]
    \centering    \caption{Best-fit stellar and accretion parameters. Uncertainties are given as 3$\sigma$ limits.}
    \label{tab::best_fit_XS}
    \begin{tabular}{l|c|c|r}
\hline
  Parameter & Value & Error & Units \\
 \hline
 SpT & M0.5 & 0.5 & ... \\
 $T_{\rm{eff}}$ & 3810 & 100 & K\\
 $A_V$ & 0 & 0.3 & mag\\
$\log(L_{\rm acc})$ & -2.092 & 0.2 & $L_\odot$\\
$\log(L_\star)$ & -0.61 & 0.15 & $L_\odot$ \\
 $M_\star$ & 0.60 & 0.05 & $M_\odot$ \\
 Age & 5.5 & 1 & Myr\\
$\log\dot{M}_{\rm acc}$ & -9.21 & 0.35 & $M_\odot$/yr\\
\hline
\end{tabular}
\end{table}

\section{The planet-forming disk}
\label{sec:disk}

Our observations detect an intricately structured planet-forming disk around 2MASSJ1612 in scattered light for the first time. In the following, we describe the SPHERE view and compare it to recently obtained ALMA submillimeter observations (\citealt{Sierra2024}). 

\begin{figure*}
\centering
    \includegraphics[width=1.0\textwidth]{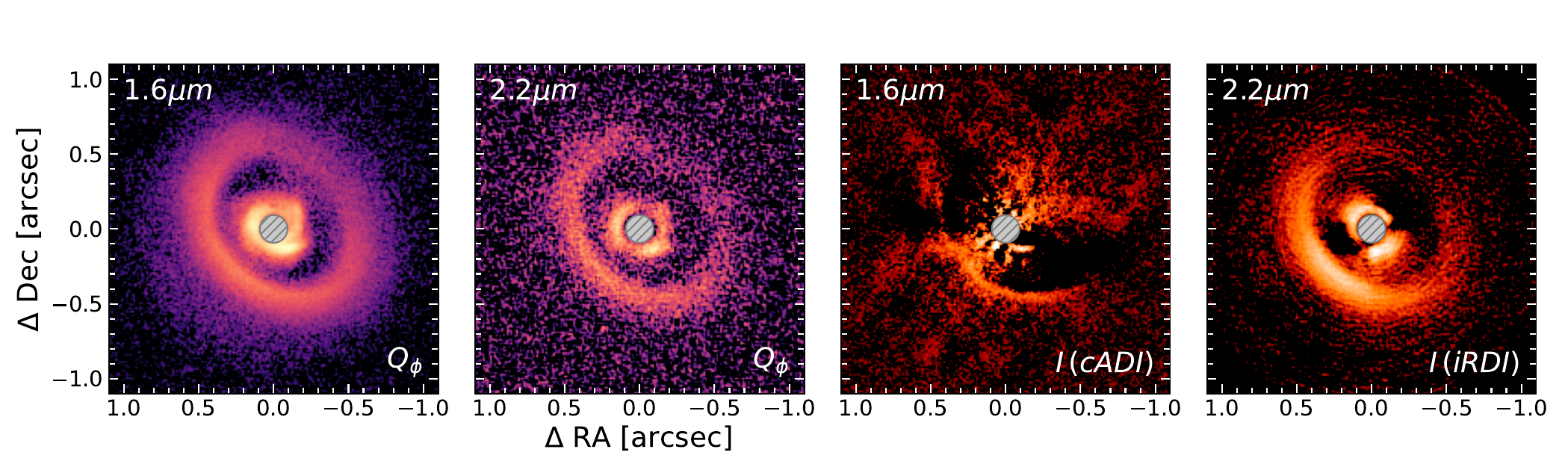}
    \caption{SPHERE observations of 2MASSJ1612 system in H band (1.6$\mu$m) and in K band (2.2$\mu$m). The two left-most images show polarized light after polarization-differential imaging, while the two right images show the total intensity after classical angular-differential imaging (cADI) and iterative reference-differential imaging. Note that the polarized light observations give a faithful image of the disk morphology, while processing artifacts are present for total intensity.}
    \label{fig::sphere_main}
\end{figure*}

\subsection{The SPHERE scattered light view}
\begin{figure}
\centering
    \includegraphics[width=0.49\textwidth]{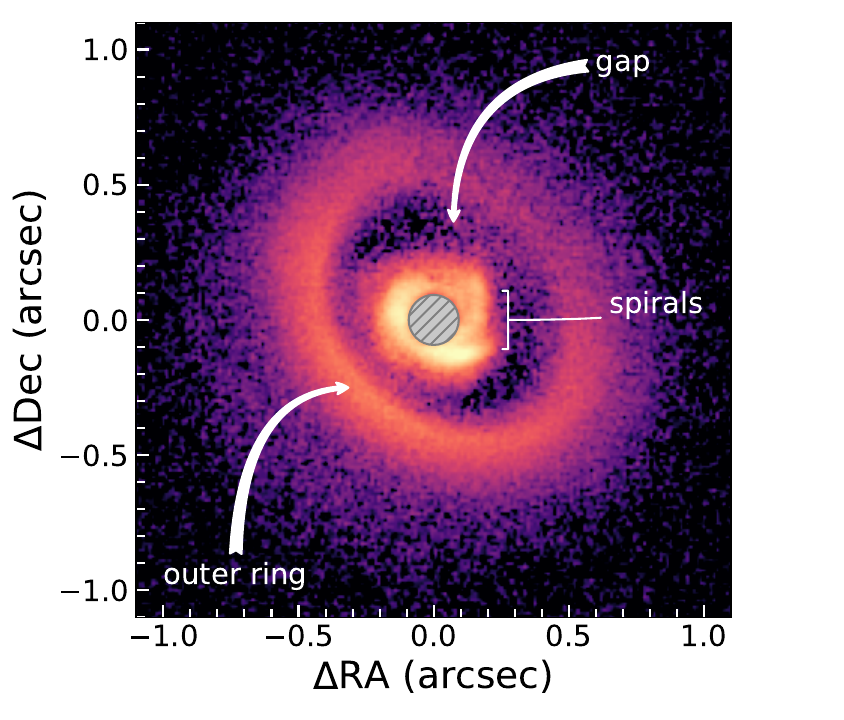}
    \caption{SPHERE H-band Q$_\phi$ image of 2MASSJ1612 with the main scattered light features annotated.}
    \label{fig::sphere_anno}
\end{figure}

The SPHERE/IRDIS polarimetric observations in Figure~\ref{fig::sphere_main} show a highly structured disk. We highlight the main visible components in Figure~\ref{fig::sphere_anno}. From the outside to the inside we see a bright outer ring at 0.6'' (along the major axis of the disk), followed by a gap in the disk with an outer edge at roughly 0.5''. We detect an inner disk with two prominent spiral arms (more visible in the H-band image), which reach out to 0.3''. The inner spirals have launching points roughly in the south and in the west of the inner disk; i.e., separated roughly 90$^\circ$ azimuthally. Using ellipse fitting (see Appendix~\ref{app::ellipse-fit}) to the outer ring \citep{Ginski2024}, we find a disk inclination of 40$^\circ \pm$2$^\circ$ and a position angle of 224$^\circ \pm$3$^\circ$, which closely resemble the values recovered from ALMA observations (i=37$^{+0.1}_{-0.2}$\,deg, p.a. = 225.1$^{+0.2}_{-0.9}$\,deg; \citealt{Sierra2024})\footnote{Note that for the position angle we added 180$^\circ$ to the value given in \cite{Sierra2024} as we followed the usual convention for scattered-light observations to encode also the position of the disk's near side in the position angle (see, e.g., \citealt{Ginski2016, Valegard2022}).}.

We find a clear offset of the outer-disk ring along the minor axis. Assuming that the disk is not strongly eccentric, we can deduce a height of the disk-scattering surface of 13.3$\pm$1.2\,au at a radial separation of 77$\pm$2\,au from this offset; i.e., an aspect ratio of 0.17$\pm$0.02 at this radial separation from the star. This makes the disk vertically thicker than the disks around T Tauri stars observed in \cite{Avenhaus2018}. According to the hydrodynamic studies by \cite{Binkert2021}, this could be an indication that there is a planet present in the disk, which leads to enhanced vertical stirring.

Using the measured disk inclination as well as the aspect ratio, we de-projected the disk to a face-on configuration in order to measure the pitch angles of the spiral arms (see Appendix~\ref{app::deprojection}). We assumed that the inner and outer disk are seen under the same inclination angle; i.e., that there is no significant misalignment between inner and outer disk. We find a pitch angle of $\sim$21$^\circ$ for the fainter arm launching roughly in the west of the disk, and a slightly smaller pitch angle of $\sim$17$^\circ$ for the brighter arm launching from the south of the inner disk. We note that the measurement for the fainter western arm might be more strongly affected by the de-projection as the arm reaches the semi-minor axis of the disk, while the majority of the arm launched from the south of the inner disk is located around the semi-major axis, which is less affected by projection effects. Thus, it may well be possible that both arms have similar pitch angles, in which case the smaller value of $\sim$17$^\circ$ may be considered more accurate. 

From the H- and K-band observations of the disk, we can obtain the disk color, which in turn is linked to the dust properties at the scattering surface (see, e.g., \citealt{Tazaki2019}). Following \cite{Avenhaus2018} and \cite{Ginski2023}, we determined the stellar flux corrected ratio of the disk integrated flux:
\begin{equation}
    H/K = \mathrm{\frac{L_{disk_H}/L_{\star_H}}{L_{disk_K}/L_{\star_K}}},
\end{equation}
wherein $L_{disk_{H/K}}$ are the integrated disk fluxes in the H and K bands in polarized light and $L_{\star_{H/k}}$ are the integrated disk and star fluxes in total intensity. The disk fluxes were measured with an annulus from eight to 120 pixels (98\,mas to 1470\,mas) in the Q$_\phi$ images shown in Figure~\ref{fig::sphere_main}. The inner eight pixels of the images were left out of the measurement as they are covered by the coronagraph. The stellar fluxes were measured in the flux calibration images attached to each sequence in which the system was imaged without coronagraph and with short integration time and a neutral density filter inserted to avoid saturation. These fluxes were then scaled to the science images given the ratio of integration times and filter throughput. We find $H/K$ = 1.5$\pm$0.1; i.e., the disk is blue between the H- and K-band filters. The blue colors are usually attributed to small grains (i.e., Rayleigh scattering). However, Rayleigh scattering would be expected to produce an equal amount of forward and back scattering and thus cannot account for the observed forward scattering in the system. A strong forward scattering with a bluish color is naturally explained if we consider large fluffy dust aggregates made of sub-micron-sized monomers (\citealt{Tazaki2019, Tazaki2023}). Such fluffy aggregates would also feature a low Stokes number and hence would be aerodynamically well supported. Thus, the presence of such dust aggregates could in principle also explain why we find a relatively high disk aspect ratio. However, we caution that the aspect ratio of the disk could also simply be explained by a higher disk gas content and thus stronger pressure support. Detailed models, tailored to the specific geometry of the 2MASS1612 system are required in the future to confirm the presence of fluffy dust aggregates in the disk.

\subsection{Detection of point sources}
\label{sec:point_source}

\begin{figure*}
\centering
    \includegraphics[width=0.99\textwidth]{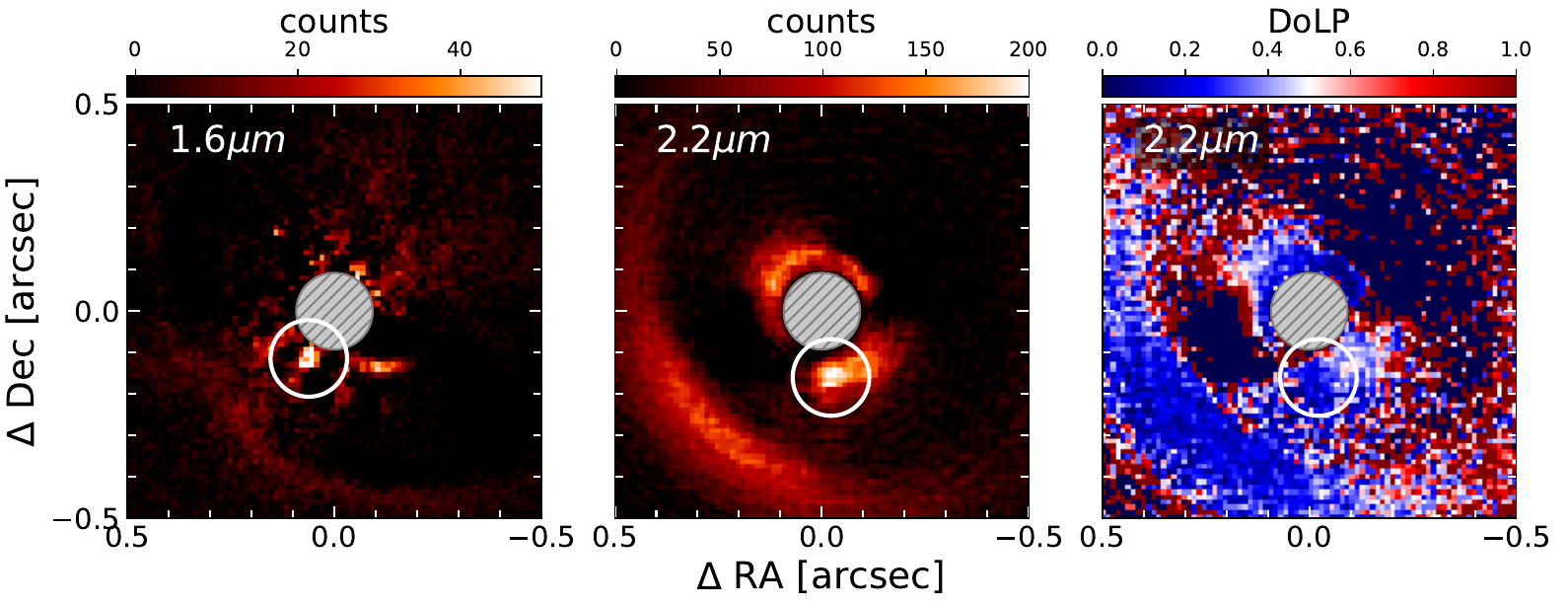}
    \caption{Total-intensity images of 2MASS1612 system; same as in Figure~\ref{fig::sphere_main}, but with a zoomed-in view of the tentative point-source detections, which are marked by a white circle. The right panel shows the degree of linear polarization for the K-band data as derived from the iRDI and polarized-intensity image. The tentative point-source location is marked and shows a lower dip than the directly adjacent inner-disk signal. }
    \label{fig::sphere_point_source}
\end{figure*}

\begin{figure}
\centering
    \includegraphics[width=0.49\textwidth]{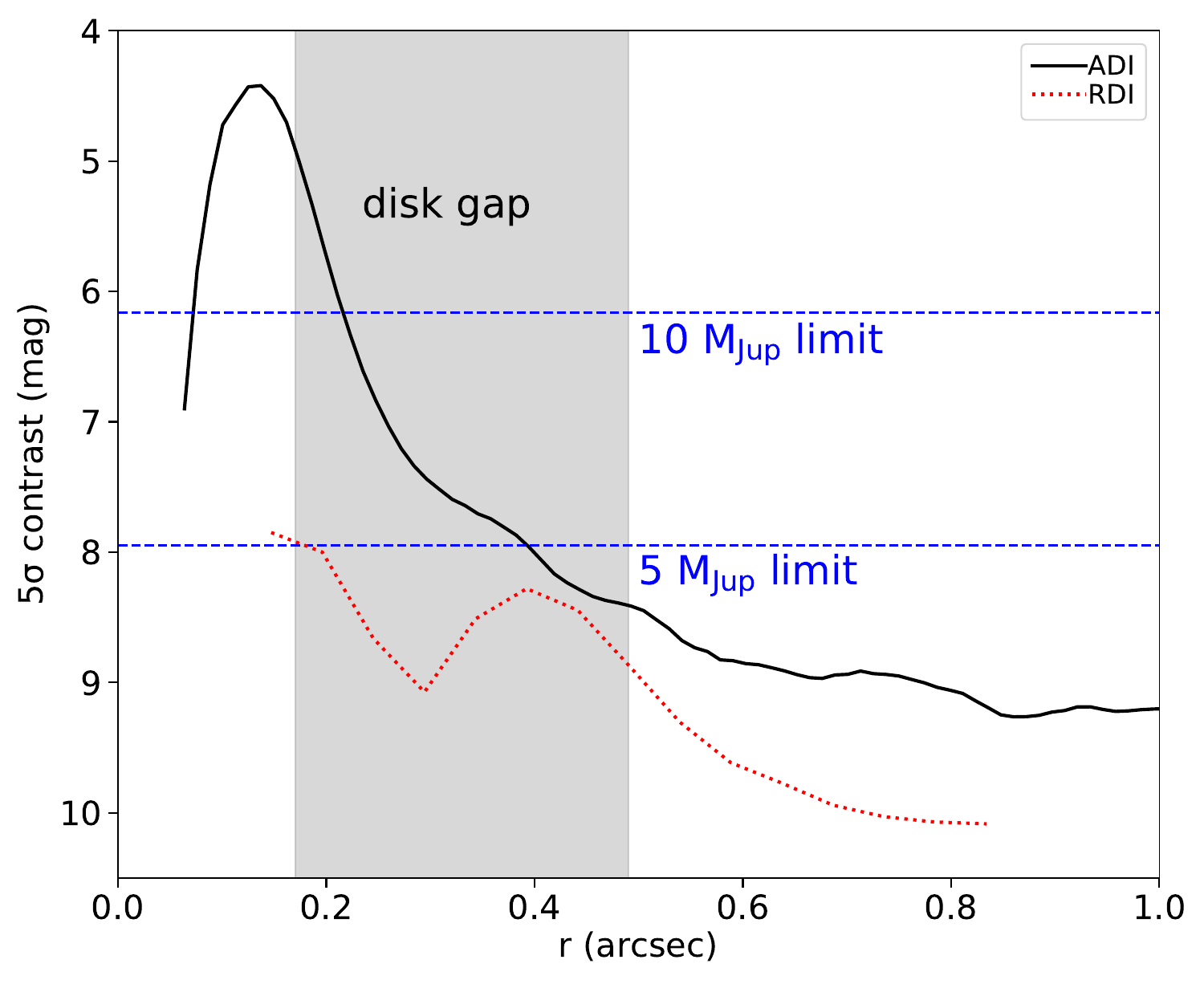}
    \caption{SPHERE K-band detection limits as function of radial separation from the central star. We indicate the approximate position of the radial gap in the disk by the gray shaded area. The solid black line gives the detection limit via fake planet injection for the ADI reduction of the K-band data. The dotted red line gives the RDI detection limit by measuring the noise in the final iterative RDI image in concentric circles and scaling the stellar signal to a signal-to-noise ratio of 5. We show corresponding mass limits via the dashed blue lines and based on {\tt AMES-DUSTY} evolutionary models.}
    \label{fig::detection_limit}
\end{figure}

In addition to the detection of the polarized scattered light signal of the planet-forming disk, the SPHERE H- and K-band observations are, in principle, also sensitive to thermally emitting point sources in total intensity.
The direct detection of an embedded proto-planet within the disk is, however, challenging, due to the possibility of false-positive signals created through the complex interplay of disk scattered-light signal with post-processing algorithms (\citealt{Milli2012,Rameau2017}) and the added extinction of possible circumplanetary material (see, e.g., \citealt{Sanchis2020}).
We are not detecting clear planet candidates inside the disk gap or outside of the disk and out to a radius of 3" ($\sim$400\,au) around the central star. While outside of the disk signal the definition of a clear detection is trivial (an unresolved source with a peak at least 5$\sigma$ above background level), it is worth briefly mentioning what we consider a clear detection in the disk gap. Such a signal should be resolved from the disk signal and should ideally be visible in both observations' epochs; i.e., in the H and K bands, with the red color consistent with a low-temperature object. This can be relaxed to a certain degree if a signal is present in the K band only, but not in the  H band, as the K-band observations are more sensitive to embedded and very red objects. If a point-source candidate is detected in the H and K bands, we would expect it to be co-located in both epochs, as orbital motion should be negligible in the short time span between observations. \\
While there is no clear detection within the disk gap, we do find two possible candidate signals in the H and K bands in close proximity to each other, but not co-located between epochs. We highlight both of these candidates in Figure~\ref{fig::sphere_point_source}. The H-band point source is located at a separation of 129$\pm$11\,mas and a position angle of 151$^\circ \pm 3^\circ$, while the K-band point source is located at a separation of 172$\pm$17\,mas and a position angle of 195$^\circ \pm 4^\circ$. The orbital separation of these point sources would be $\sim$22\,au if they were in the disk plane. This means we would only expect a motion of $\sim$0.5$^\circ$ within the time span of two months between the two observation epochs (i.e., significantly less than what is observed). Therefore, it appears unlikely that these two tentative point sources trace the same signal, and it is indeed possible that they are data-processing artifacts. This is particularly true for the H-band candidate source, which has been found using classical ADI processing, which in turn is particularly prone to creating artifacts from disk signal. The K-band candidate source on the other hand was found in the iRDI processed image. Although RDI can still lead to artifacts due to over-subtraction, this should be mitigated by the iterative iRDI process (see \citealt{Ginski2021} and \citealt{Stapper2022}).  \\
In the case of the K-band candidate source, we do not have a clear separation between the potential point source and the inner-disk signal, which makes the detection somewhat uncertain. 
Further analysis of this tentative point source involved fitting a 2D Gaussian, while a two-degree 2D polynomial was used to model the large-scale emission from the surrounding disk, ensuring the Gaussian fit isolated the point source. The fit was performed using the Levenberg-Marquardt least-squares optimization method; this yielded a resulting FWMH of 0.0621 $\pm$ 0.0297\,arcsec. Within this uncertainty the FWHM is consistent with the expected telescope resolution in the K band of 0.0675\,arcsec. This suggests the source may be unresolved, which is consistent with the characteristics of a compact point source. However, the relatively large uncertainty ($\pm$ 0.0297) leaves some ambiguity.
We note that at the position of the K-band point source we do detect a drop in polarized light signal in the K band, which is in principle consistent with thermal radiation from an embedded proto-planet (see Figure~\ref{fig::sphere_point_source}, right panel). This was indeed one of the criteria used to argue that the candidate source in the AB\,Aur system traces planet thermal emission (\citealt{Currie2022}).
We measure a K-band magnitude contrast of $\sim$9.2\,mag relatively to the central star for this candidate source. Using {\tt AMES-DUSTY} atmosphere models (\citealt{Allard2012}) for an age of 5\,Myr, this translates into a mass of 4\,M$_\mathrm{Jup}$ \footnote{If we assume an older age of 10\,Myr in line with the typical UpperSco estimate, the mass would be slightly higher; i.e., $\sim$5\,M$_\mathrm{Jup.}$}. This estimate assumes that we detect predominantly thermal emission at the source location, which is not significantly attenuated by surrounding disk material.\\
We used our K-band data, which should be more sensitive to intrinsically red low-temperature objects and embedded planets to determine general detection limits in the K-band star-hopping data set. We did so in two different ways using the ADI and iterative RDI reductions of the data set. To calculate the ADI contrast curve, we made use of the Vortex Image Processing package (\texttt{VIP})\footnote{https://vip.readthedocs.io/en/latest/index.html.} package \citep{2017AJ....154....7G,2023JOSS....8.4774C}, which calculates the final image using a principal component analysis (PCA), and injected fake planets to better evaluate the contrast curves. For the iterative RDI data set, self-subtraction is not a problem. Thus, we directly measured the noise in concentric annuli. Using the flux of the star, retrieved from the flux-calibration observations, we then calculated the 5$\sigma$ detection threshold for each annulus. 
The results are shown in Figure~\ref {fig::detection_limit}. 
As discussed by \cite{Wahhaj2021}, for example, the RDI reduction produces significantly deeper detection limits at small separations from the central star due to the lack of signal self-subtraction compared to the ADI. However, we caution that the determination of the noise background in the image in the presence of disk signal likely overestimates the noise (since disk signal is unavoidably polluting the measurement annuli). Thus, in principle slightly deeper detection limits than those shown in Figure~\ref{fig::detection_limit} are possible, as evidenced by the detection of the tentative point source in the disk gap in the K-band data. Keeping this in mind, we find that inside the gap we are roughly sensitive to planets more massive than $\sim$5\,M$_\mathrm{Jup}$ depending on the radial distance from the central star. Outside of the outer disk ring, we reach mass limits of roughly 3-4\,M$_\mathrm{Jup}$. These mass limits were computed from the contrast limits (also presented in Figure~\ref{fig::detection_limit}) by using the {\tt AMES-DUSTY} atmosphere models (\citealt{Allard2012}). Based on our XSHOOTER analysis, we assumed a system age of 5\,Myr. We further assumed that the disk gap is mostly cleared of material; i.e., we do not account for possible embedding of the planets. If there is still a significant amount of dust in the gap, our mass limits are too optimistic.

\subsection{Comparison with ALMA Observations}
\begin{figure}
\centering
    \includegraphics[width=0.49\textwidth]{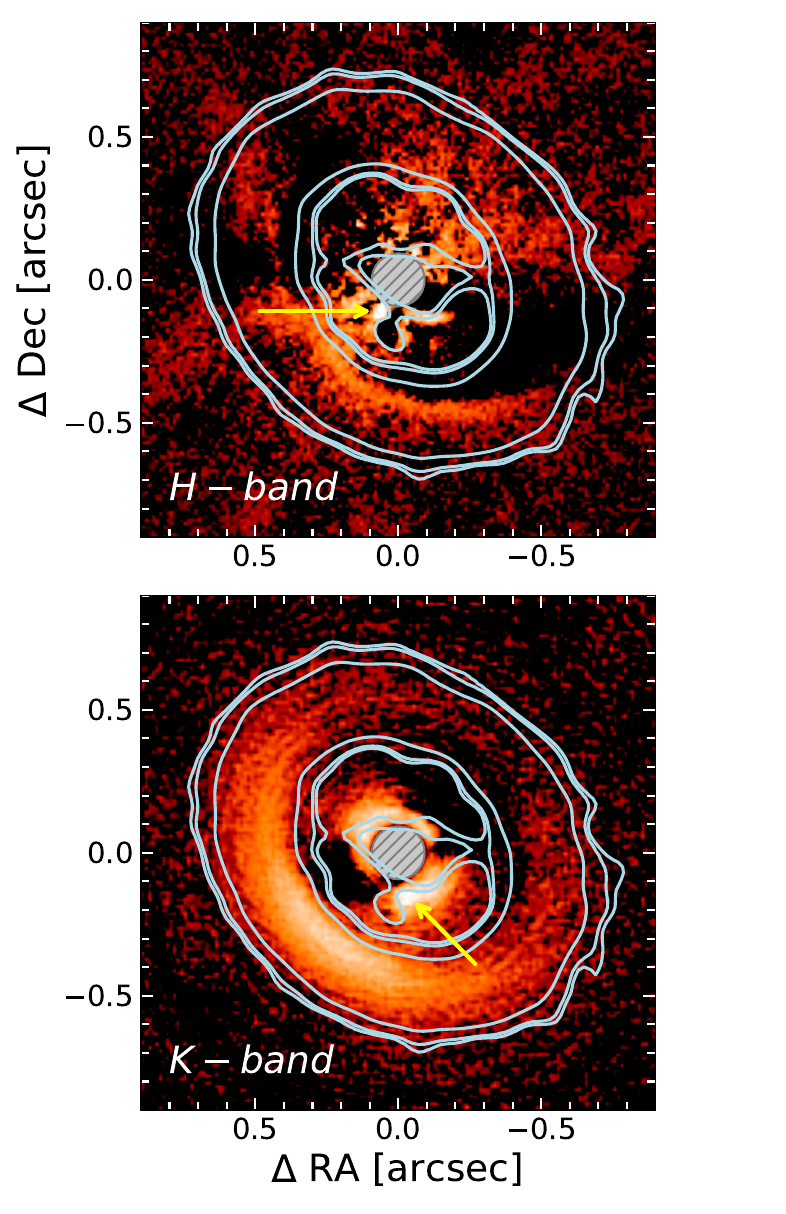}
    \caption{SPHERE total-intensity reduction of the H- and K-band data sets (with ADI and iRDI as in the previous figure). Total intensity observations are sensitive to thermal emission of embedded planets. We show the ALMA continuum data from \cite{Sierra2024} as the contour
overlay (3$\sigma$,4$\sigma,$ and 10$\sigma$ contours shown). The yellow arrows indicate tentative point sources detected with SPHERE in the H and K bands. Both detections are directly adjacent to a 3$\sigma$ compact continuum signal detected with ALMA in the disk gap.}
    \label{fig::alma-overlay}
\end{figure}

As part of the ALMA Large Program AGE-PRO: ALMA survey of Gas Evolution in Protoplanetary disks, 2MASSJ1612 was observed in Band 6 (1.3\,mm) in addition to observations of $^{12}$CO (J=2-1), $^{13}$CO (J=2-1), C$^{18}$O (J=2-1), and H$_2$CO (J=$3_{(0,3)}-2_{(0,2)}$) (\citealt{Sierra2024}). The dust continuum emission was observed at a resolution of 0.17''$\times$0.14''. These ALMA observations also resolved several structures, including a cavity and a ring-like structure that peaks at 0.57'' ($\sim$75\,au). There is the detection of a faint inner disk in the ALMA observations that remains unresolved. The minimum level of the gap is located at 0.24'' ($\sim31$\,au).  The cavity is also detected in the molecular lines, except in the $^{12}$CO, which could be due to the high optical depth of this line. An additional compact dust continuum source inside the cavity at 0.24'', with an estimated dust mass between 1.8 and 8.4 lunar masses was detected at the 3$\sigma$ level. This compact emission has been proposed as a circumplanetary disk, similar to the one observed around PDS\,70c \citep{benisty2021}. Moreover, \cite{Sierra2024} suggested the presence of spiral arms in the residual maps when modeling the dust continuum. Figure~\ref{fig::alma-overlay} shows an overlap of the SPHERE total intensity observations in the H and K bands (in colors) and ALMA observations (in contours). Several of the detected substructures are observed in both scattered light and submillimeter continuum emission, including the inner disk, the gap, and the ring. The inner spiral arms observed with SPHERE are not detected with ALMA. Remarkably, both of the candidate point sources in the SPHERE H- and K-band observations (as highlighted with a yellow arrow in Figure~\ref{fig::alma-overlay}) are very close to the location of the compact emission (potentially a circumplanetary disk) detected with ALMA.  Deeper and higher angular resolution observations are needed to better understand the nature of such compact emission, the point sources, and their potential connection.

\section{Discussion}  \label{section:discussion}
\label{sec:discussion}

\subsection{Signatures of planet-disk interaction}

\begin{figure*}
\centering
    \includegraphics[width=1.0\textwidth]{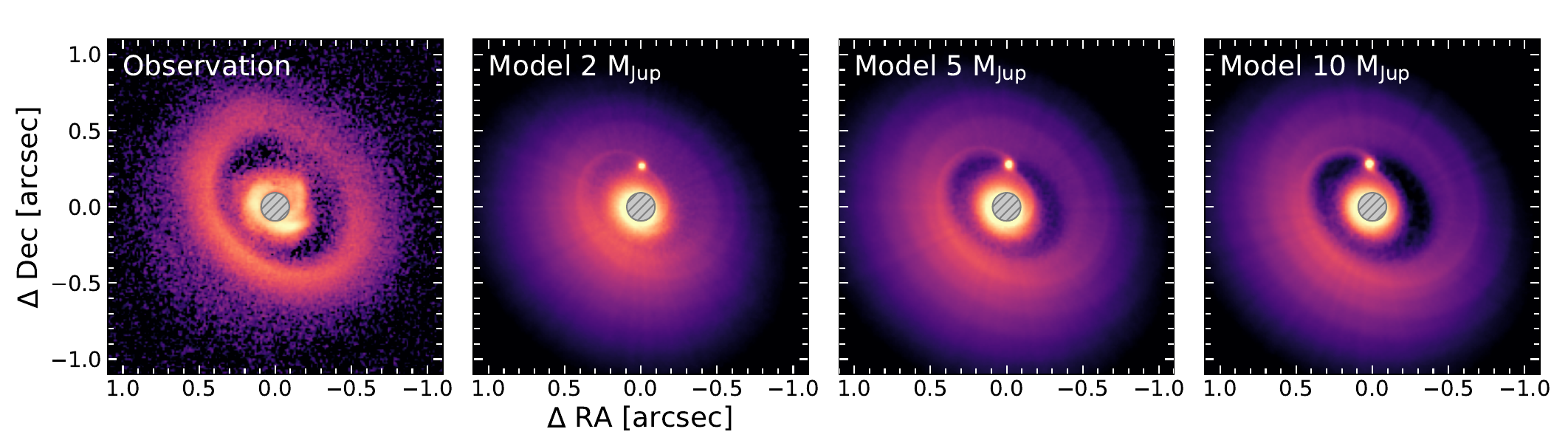}
    \caption{SPHERE observations of 2MASSJ1612 system in H band (1.6$\mu$m) and radiative-transfer-model images created from hydrodynamic simulations. The only parameter varying between the models is the mass of the injected planet indicated in the upper left of each panel. Note that we show total intensity, including the planet thermal emission for the models, while the polarized-light image of the observation is not sensitive to planet thermal emission.}
    \label{fig:models}
\end{figure*}

The signatures that are visible in scattered light show a strong similarity to hydrodynamic simulations of planets embedded in disks (e.g., \citet{Dong2016, Bae2016, Zhang2018}). To directly compare our observations to planet-induced disk structures, we ran dedicated hydrodynamic simulations with the code {\tt PHANTOM} \citep{Price2018_Phantom} and then post-processed these observations with the radiative transfer code {\tt MCFOST} \citep{Pinte2006, Pinte2009} to create simulated observations. We describe the details of the simulation in Appendix~\ref{app::hydro}. The resulting images are shown in Figure~\ref{fig:models}. We find that perturbing planet masses above $\sim$5\,M$_\mathrm{Jup}$ reproduce the observed features in the disk, such as the clear gap and the opening angles of the planet-driven spiral arms assuming a single perturber. We note that the gap can, in principle, also be opened by multiple lower mass planets. However, for lower planet masses we would expect more tightly wound spirals \citep[see, e.g., the grid studies in ][]{Zhang2018}.

To further narrow down the mass range of perturbing planets, we compare the direct measurements of disk-gap width, spiral-arm contrast, and spiral-arm pitch angle as well as grain segregation at the gap edge to hydrodynamic and radiative-transfer simulations from the literature.

Following \citet{Kanagawa2016}, we can estimate the perturbing planet mass from the width of the gap the planet carves in the disk. Assuming that the disk's vertical height profile follows a power law with an exponent of 1.22 \citep{Avenhaus2018} and given our height measurement of $\sim$13\,au at the ring location of 77\,au, we expect an aspect ratio of $\sim$0.15 in the center of the disk gap at $\sim$40\,au\footnote{This is the aspect ratio for the scattering surface. Following \citet{Chiang2001}, we reduced this by a factor of four for the gas scale height. We note that this is the aspect ratio prior to opening the gap by the putative planet; i.e., the actual aspect ratio of the disk inside the gap should be considerably lower at the present time.}. The gap width that we measure is $\sim$46\,au along the disk's major axis. We assume a disk viscosity $\alpha$ parameter \citep{Shakura1976} of 10$^{-3}$. Using the formula provided in \citet{Kanagawa2016}, we then find a planet to star mass ratio of 1.8$\times$10$^{-3}$; i.e., a planet mass of 1.1\,M$_\mathrm{Jup}$. If we vary the viscosity parameter between 10$^{-5}$ and 10$^{-2}$,  it leads to a mass range between 0.1\,M$_\mathrm{Jup}$ and 3.4\,M$_\mathrm{Jup.}$  If we instead keep the viscosity parameter fixed at the nominal value of 10$^{-3}$ but allow for a slightly smaller discrepancy between the gas scale height and the height of the scattering surface of a factor of three (see \citealt{Chiang2001}), we obtain a planet-mass estimate of 1.7\,M$_\mathrm{Jup}$. Finally, if we allow for a larger stellar mass of 0.7\,M$_\odot$ as indicated by the ALMA CO velocity fit, all estimated planet masses increase by a factor of 1.2. Considering the uncertainties in the assumed parameters, this yields a range of planet masses between 0.1\,M$_\mathrm{Jup}$ and 4.1\,M$_\mathrm{Jup}$ when applying the hydrodynamic results of \cite{Kanagawa2016}.\\ 
To understand the model dependency of these results, we also estimated the perturbing planet mass using the relations and models derived by \citet{Zhang2018}. Using the relation of the gap-width $\Delta$ and their fitting parameter K', $K' = A\Delta^B$, where $\Delta$ is defined as $\frac{r_{out} - r_{in}}{r_{out}}$, we calculated K' based on our measured gap width of 46 au. Assuming an outer gap radius of 77 au and using the best-fit gas surface-density parameters from Table 1 of \citet{Zhang2018}, we obtained a K' value of 0.11. Applying this K' value to the empirical relation from the same equation,
\begin{equation}
    \frac{K'}{0.014} = \frac{q}{0.001}\left(\frac{\frac{h}{r}}{0.07}\right)^{-0.18}\left(\frac{\alpha}{10^{-3}}\right)^{-0.31}
,\end{equation}
where $q=\frac{M_p}{M_*}$, we find a planet-mass value of 4.7\,M$_\mathrm{Jup}$. To account for potential uncertainties in determining the exact position of the outer edge of the gap, we repeated this calculation for outer radii of 75\,au and 80\,au, yielding planet masses of 5.2\,M$_\mathrm{Jup}$ and 4.1\,M$_\mathrm{Jup,}$ respectively. This trend, where decreasing $r_{out}$ leads to an increasing planet mass, is consistent with the $\Delta$ versus K' fits seen in the gas panel of Figure 12 of \citet{Zhang2018}. Specifically, for a disk with $\frac{h}{r} = 0.05$ (similar to our value of 0.04-0.05 considering the factor of 3-4 between scattering height and gas scale height) and viscosity parameter $\alpha = 10^{-3}$, $\Delta$ values between 0.5 and 0.6 correspond to planet masses of $\sim$ 5\,M$_\mathrm{Jup}$. This is slightly higher than the upper end of the mass range derived from the \cite{Kanagawa2016} simulations.

We also used the spiral-arm properties to determine the mass of the possible perturbing planet.
We compared the contrast of the spiral arms relative to the surrounding disk signal and the spiral-arm pitch angle to simulations by \citet{Dong2015}. That work found that a planet with a mass ratio of 6$\times$10$^{-3}$, i.e. 4\,M$_\mathrm{Jup}$ in this case, produced spiral arms that have a brightness enhancement of 150\,\% relatively to the surrounding disk and pitch angle between 10$^\circ$ and 16$^\circ$. For the brighter spiral arm, launching from the south of the inner disk, we find a brightness enhancement of 170\,\% and a pitch angle of 17$^\circ$, broadly consistent with their simulations. However, the brightness enhancement in \citet{Dong2015} was computed for a face-on disk, while in our case the disk is significantly inclined. This may lead to an additional, non-perturber-related, brightness contrast due to different light-scattering angles at both spiral-arm locations, which could artificially enhance the contrast. This might imply a lower planet mass than 4\,M$_\mathrm{Jup}$ for the case of the potential planet driving the spirals in 2MASS1612.

By comparing the cavity size at the two wavelengths, i.e., in near-infrared scattered light and millimeter emission, it is possible to give an estimation of the planet mass creating the gap or cavity as proposed by \citet{ovelar2013}. This difference in the gap size is a natural result of dust trapping at the outer edge of planet-induced gaps. Although several other mechanisms can explain the formation of large cavities in disks, such as dead zones \citep[regions of low disk ionization,][]{flock2015}, large radial segregation between the distribution of small (micron-sized) and large (millimeter-sized) grains is a unique characteristic seen in models of dust trapping by planets \citep{pinilla2017}.
The gap size in the SPHERE observations is roughly 0.5", as measured along the disk's major axis, while in ALMA the peak of the millimeter emission is at 0.57''. Using the ratio of these two values, \citet{ovelar2013} provided a fitting function for obtaining the planet mass. In the case of 2MASSJ1612, this ratio points to a planet mass between 1-2\,$M_{\rm{Jup}}$ located between 0.2 and 0.3''.

The different substructures observed in the disk all point toward the presence of a massive gas giant in the observed disk gap. Depending on the utilized model and their underlying assumptions, we find an approximate mass range between 0.1\,$M_{\rm{Jup}}$ and 5\,$M_{\rm{Jup}}$ for this planet. Thus, while the different models include (or exclude) different physical processes in the disk and make slightly different assumptions on the disk structure, they together yield a consistent picture.
Given our mass limits, a non-detection of the planet in the SPHERE K-band observations is consistent with the predicted planet mass range from the simulations. 

\subsection{Other explanations for substructures}

While the structures observed in the disk of the 2MASS1612 system appear to be consistent with a planet-disk interaction hypothesis, there are other possible scenarios that can create a similar disk morphology. In the following, we discuss several possible ways to explain the observed disk structures without invoking the presence of a perturbing planet. As disk gaps or cavities can be created by a multitude of processes (see, e.g., the recent review by \citealt{Bae2023}), we focus the discussion on processes that can also generate the spiral arms in the inner disk.\\

\subsubsection{Disk structures excited by a stellar companion}

If 2MASS1612 is not a single star but has a close stellar binary companion, this may explain some of the disk structures. In particular stellar companions can drive spiral density waves inside and outside their orbital position (see, e.g., \citealt{Goldreich1979}). Given the detection limits from our SPHERE observations presented in Figure~\ref{fig::detection_limit}, we can rule out the presence of stellar companions down to $\sim$0.1\,arcsec (13\,au). This maximum angular separation would place a potential stellar companion  inside the spiral-arm features and would make the inner disk observed in scattered light a circumbinary disk. Using our XSHOOTER observations we can place some limits on the presence of close, double-lined spectroscopic binaries. Given the spectral resolution in the optical wavelength range we would have been able to resolve Doppler-shifted spectral lines of a stellar companion with an (inclination-dependent) orbital velocity of $v\,sin(i) = 16\,km\,s^{-1}$. Under the assumption that such a stellar companion might be co-planar with the disk, this gives an effective orbital velocity of $25\,km\,s^{-1}$. A stellar companion should not be more massive than the primary star, otherwise it would dominate the stellar spectrum. If we assume an equal-mass binary, the spectral lines of such an object would have been detected within our XSHOOTER spectra up to a maximum orbital separation of 1\,au. If the binary companion were instead a low-mass M-type star with a mass of 0.1\,M$_\odot$, this maximum orbital separation would shrink to 0.1\,au. In principle, this leaves an orbital parameter space between 0.1\,au and 13\,au for which we cannot directly rule out the presence of a binary companion. However, at such a close orbital separation one would expect that the companion orbital motion introduces an astrometric signal, which may have been detected by Gaia. We thus inspected the renormalized unit weight error (RUWE) of the Gaia astrometric solution for 2MASS1612. We found a RUWE of 1.3, which is slightly larger than the nominal value of $\sim$1 for a single-star solution. However, as recently found by \cite{Fitton2022}, the presence of circumstellar disks inflates the RUWE values, and they recommend a cutoff of 2.5 for such systems and acceptable single-star solutions. As the RUWE for 2MASS1612 is well within this range, we can conclude that it is consistent with a single-star system. Thus, the body of observations collected for the system so far does not indicate the presence of a close stellar companion. Furthermore, while such a companion can drive spiral arms outside of its orbit, these spiral arms are typically located at the outer edge of a cavity created by companion-disk interaction (see, e.g., simulations by \citealt{Price2018} and \citealt{Calcino2023}). This morphology appears inconsistent with the inner disk signal that we detected all the way down to the coronagraphic mask inside the spiral features.
We thus conclude that a stellar companion is unlikely to be the cause of the observed disk features.\\

\subsubsection{Disk gravitational instability}

\begin{figure}
\centering
    \includegraphics[width=0.49\textwidth]{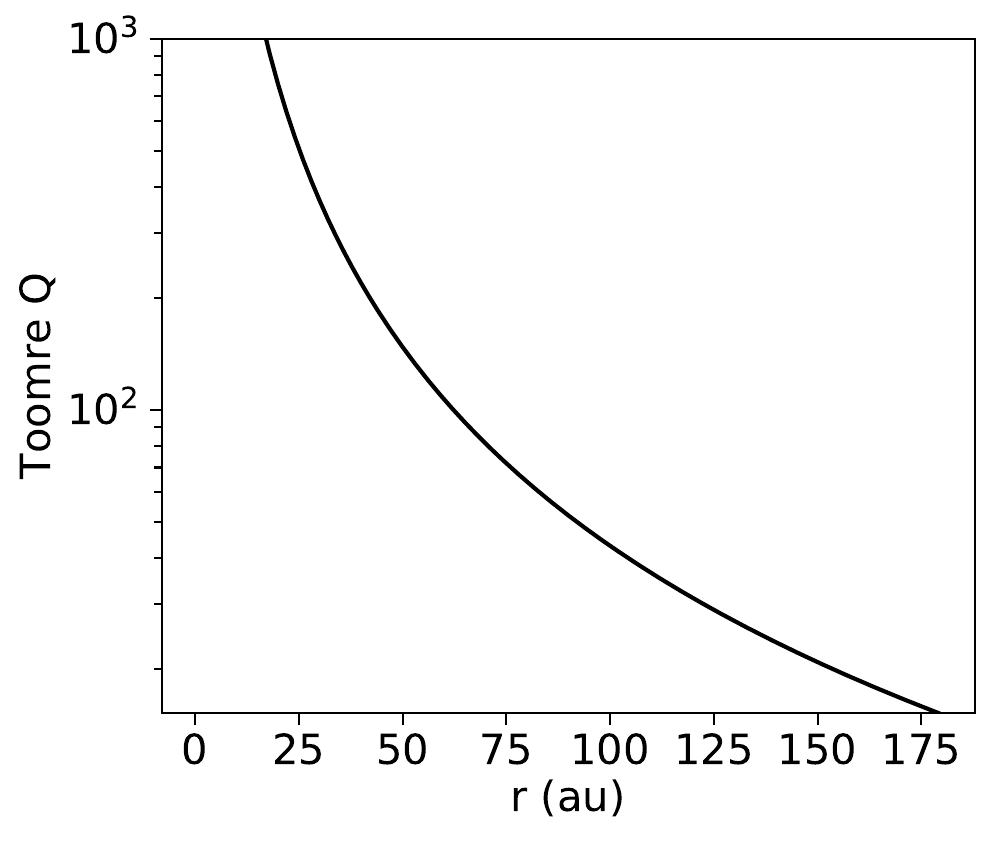}
    \caption{Estimated (dimensionless) Toomre Q parameter for disk stability plotted against a range of disk radii using the measured disk aspect ratio and the ALMA disk mass from \cite{Sierra2024} as input. The disk appears stable against gravitational collapse, as Q always remains larger than 15.}
    \label{fig::toomreparameter}
\end{figure}

Another candidate process for the formation of the inner spiral arms could be self-gravitating waves generated by gravitational instabilities in the disk (\citealt{Cossins2009, Dong_2016SpiralInstability, Baehr_2021}). To obtain a back-of-the-envelope calculation for whether 2MASS1612 is susceptible to gravitational collapse, the Toomre Q-parameter is invoked: 
\begin{equation}
Q = \frac{c_s\Omega}{\pi G \Sigma},    
\end{equation}
where $c_s$ is the temperature-dependent sound speed in the disk, $\Omega$ is the Keplerian orbital frequency, $\Sigma$ is the surface density, and $G$ is the gravitational constant.  
Following \cite{Durisen2007}, a $Q \leq 1.7$ would indicate that the disk is susceptible to non-axisymmetric perturbations of the kind that could give rise to the distribution of spiral arms present in the data.
Using the relation between the disk's gas-pressure scale height, $h_g,$ and sound speed,
\begin{equation}
h_g = \frac{c_s}{\Omega} 
,\end{equation}
as well as the formula for the Keplerian rotational frequency,
\begin{equation}
\Omega = \sqrt{GM_*/R^3}
,\end{equation}
we can express the Toomre $Q$ parameter as
\begin{equation}
Q=\frac{h_g M_*}{\pi \Sigma R^3}.
\end{equation}
We can approximate the gas-pressure scale height using our measurement of the disk's aspect ratio and assuming a flaring disk profile of the following form:
\begin{equation}
h_g(R) = h_{g,0} (R/R_0)^\beta.
\end{equation}
For $h_{g,0,}$ we assume a value of 13.3\,au as measured from our scattered-light image, divided by a factor of four to compensate for the difference between the near-infrared $\tau=1$ scattering height and the gas scale height. This height value, $h_{g,0,}$ is measured at a disk radial location, $R_0,$ of 77\,au. For the flaring exponent, $\beta,$ we assume the average value of 1.22 that was found by \cite{Avenhaus2018} for several T Tauri stars. Using the disk's height profile, we can express the Toomre $Q$-Parameter as
\begin{equation}
Q=\frac{h_{g,0}M_*}{\pi \Sigma R^3} \biggl( \frac{R}{R_0} \biggr)^\beta.
\end{equation}
To estimate the disk's surface density, we took the disk's dust mass from \cite{Sierra2024} and converted it into a total disk mass assuming a gas-to-dust ratio of 100. This yields a disk mass of 850\,M$_\oplus$.
For this simple estimation, we assumed that the disk mass is evenly distributed over the entire surface of the disk using a disk outer radius of 185\,au based on the gas-disk outer radius from \cite{Sierra2024}. Of course, the surface density will be radial dependent in a more realistic model. However, with this simple prescription, we overestimate the surface density in the outer disk, which is most susceptible to gravitational instability. Thus, if we find with this simple prescription that the disk is not likely to be gravitationally unstable, this result will not change with a more complex parameterization of the surface density.\\
We show the final result of this calculation in Figure~\ref{fig::toomreparameter} as a function of radial position within the disk. We find that with the discussed assumptions the disk is not likely to be gravitationally unstable, with $Q$ remaining as large as 15 at 180\,au. If we allow for a disk mass that is a factor of ten larger -- for example due to an extreme gas-to-dust ratio or due to self-shielding of the dust millimeter emission in the ALMA data -- $Q$ enters the instability regime outside of $\sim165$\,au; hence, even in this case it is unlikely that the disk instability could be responsible for the formation of the spiral arms in the inner-disk region. We also investigated if $Q$ enters the instability regime if we assume a constant disk aspect ratio (i.e., the flaring exponent $\beta =1$). This does not change the results significantly. We thus conclude that while our estimate for $Q$ depends on several assumptions, overall it is unlikely that the disk is gravitationally unstable.\\

\subsubsection{Magneto-hydrodynamic effects}

Magneto-hydrodynamic (MHD) effects can cause spiral structures in the disk. In particular, \cite{Heinemann2009} showed that these can be driven by magneto-rotational instability (MRI). However, as found in \cite{Flock2011}, for example, and discussed in \cite{Bae2023}, spiral density waves forming two dominant spirals are rare in this scenario. Instead, higher order spiral modes are typically excited. Thus, while we cannot rule this effect out for the 2MASS1612 system, the disk morphology would appear to be somewhat atypical for such an excitation mechanism.

\subsection{Similarities with other systems}

While either rings or spiral substructures have been found in a significant portion of planet-forming disks imaged in near-infrared scattered light \citep[see, e.g., ][ for a recent review]{Benisty2023}, the combination of both features is rare.
At the time of the publication of this study there were 33 systems known in the literature in which spiral features have been detected as disk substructures with scattered-light observations. Of these, the majority (21 systems) were recently summarized in \cite{Benisty2023}. To these we add the following systems, for which observations were published after that review article: EM$^*$\,SR24, As\,205, FU\,Ori (\citealt{Weber2023a}); V960\,Mon (\citealt{Weber2023b}); S\,CrA, CHX\,22 (\citealt{Zhang2023}); HD\,169142 (\citealt{Hammond2023}); PDS\,70 (\citealt{Juillard2022, Christiaens2024});
GM\,Aur, UY\,Aur (\citealt{Garufi2024}); V599\,Ori, V351\,Ori (\citealt{Valegard2024}). 
Of these 33 systems with detected spiral structures, only 11 show additional ring-like substructures: SR21, HD\,169142, PDS\,70, HD\,34700, S\,CrA, AT\,Pyx, HD\,100453, HD\,142527, GG\,Tau, AB\,Aur, HD\,135344\,B. These systems can be roughly divided in two categories based on whether the radial launching point of the spiral features is inside or outside the ring structure.\footnote{There are many other categorizations possible of course; for example based on the morphology and symmetry of the spiral arms (see, e.g., \cite{Benisty2023}).} Of the mentioned systems, roughly half have spiral features inside a disk ring or cavity: SR\,21, HD\,169142, S\,CrA, AT\,Pyx, GG\,Tau, AB\,Aur.
Of these six systems, S\,CrA and GG\,Tau show a complex morphology that is likely created by the existing binary companions in the systems (\citealt{Zhang2023, Keppler2020}). AT\,Pyx does not have a confirmed stellar companion; however, the disk structure is likewise complex, with possibly three spiral arms in the inner disk and an asymmetric structure outside that region that may trace either a ring or possible vortex-like features (\citealt{Ginski2022}).

 The system that most closely resembles 2MASSJ1612 is the SR\,21 system \citep{Muro-Arena2020}. Both systems show prominent double-armed spirals in the inner disk, separated with a gap from an outer ring. In both cases the spiral arms are located within a large millimeter emission cavity.
While the disk substructure of both systems closely resemble each other, the central stars differ significantly in mass, with SR21 being an intermediate-mass star of spectral type F7, while our analysis of 2MASS1612 suggests a spectral type of M0.5. Both systems are, however, in the older age range for a gas-rich disk to still be present, with 10\,Myr and 5\,Myr, respectively. This further strengthens the picture suggested in \cite{Garufi2018}, which find that spiral arms are predominantly found in older planet-forming disks.\footnote{Although, we note that the precise age determination of young intermediate-mass stars is challenging (see, e.g., \citealt{Soderblom2010}).}

Moreover, 2MASS1612 shares very interesting similarities with the planet-hosting PDS\,70 system \citep{Keppler2018, Haffert2019}. The stellar mass for PDS\,70 is $\sim 0.8\,M_\odot,$ while for 2MASS1612 it is $\sim 0.6\,M_\odot$, both of them being of a similar age (5.4\,Myr for PDS\,70 and 5-10\,Myr for 2MASS1612) and having low accretion rates \citep[$\log \dot{M}_{\rm acc}=-10.06 \pm 0.11$ for PDS\,70][]{Campbell-White2023}. 
Both systems show a visible gap in the scattered-light disk on similar scales (45\,au for PDS\,70 and 40 to 65\,au for 2MASS1612). Furthermore, the PDS\,70 system shows a similar location of peak millimeter emission with ALMA (75\,au for both systems) for the outer ring and an inner-disk emission, as is the case of 2MASS1612 (\citealt{Sierra2024}).  The major differences between the two systems are that in PDS\,70 no scattered-light spirals in the inner disk are observed. Also, the difference in position between the scattered-light disk wall locations relatively to the ALMA millimeter-peak emission between the two systems suggest less severe dust segregation in 2MASS1612. This may indicate that the planet mass in the system is lower than the mass of the outer planet in the PDS\,70 system, i.e., PDS\,70\,c (mass range between 1\,M$_\mathrm{Jup}$ and 10\,M$_\mathrm{Jup}$,  \citealt{Bae2019, Stolker2020, Wang2021}). The absence of spirals in the inner disk of PDS\,70 compared to 2MASS1612 may point to an overall lower viscosity in the disk around PDS\,70. It is visible in the gas surface-density models by \cite{Zhang2018} that for the same perturber mass and disk structure, spiral arms appear more prominently in high-viscosity disks. This would also fit with the fact that less dust segregation is observed in 2MASS1612 compared to PDS\,70, since higher $\alpha$ values of $\sim$10$^{-2}$ lead to more dust diffusion and less radial difference between the distribution of small and large grains \citep{ovelar2016}. 

Finally, 2MASS1612 also shares some similarities with the recently detected proto-planet candidate in the HD\,169142 system (\citealt{Gratton2019,Hammond2023}). Both systems show a prominent gap and outer ring as well as a bright inner disk in scattered light. \cite{Hammond2023} showed that the planet candidate appears to drive a spiral wave in the outer disk. However, the inner disk in HD\,169142 appears unperturbed and ring-like, compared to the prominent spirals in the 2MASS1612 system. The detection of an outer spiral arm but no inner spirals in this case could possibly be related to the location of the planet candidate close to the edge of the outer disk at 37\,au in HD\,169142. If on the other hand the tentative point sources that we discussed in Section~\ref{sec:point_source} indeed trace thermal emission of an embedded planet, it is located close to the edge of the inner disk, possibly making the inner-disk spirals more prominent than their outer-disk counterparts. 

\section{Summary and conclusion}
\label{sec:summary}

We present near-infrared imaging and UV-to-near-infrared spectroscopic observations of the 2MASS1612 system for the first time. The imaging observations reveal a gas-supported, potentially planet-forming disk in polarized scattered light. Remarkably, the disk shows spirals as well as a ringed substructure, which is rare among the sample of disks resolved in similar observations. We discuss several mechanisms that can create spiral substructures in a disk and find that in the case of 2MASS1612 these structures are most consistent with the presence of an embedded giant planet interacting with the disk. Our hydrodynamic and radiative-transfer models suggest a lower mass limit of $\sim$5\,M$_\mathrm{Jup}$ to open a gap similar to the one that we observe. Comparing the spiral-arm properties in the inner disk, as well as the gap width and grain segregation between near-infrared and millimeter-wavelength observations to literature models, yields a broadly similar mass range between 0.1\,M$_\mathrm{Jup}$ and 5\,M$_\mathrm{Jup}$ for the perturbing planet. This is consistent with a non-detection of the thermal radiation of this planet in our SPHERE observations. We note that we find two tentative candidate point sources at close separation in H- and K-band observation epochs that are in direct proximity to a compact continuum-emission source located with ALMA within the disk gap. If the ALMA signal is indeed related to circumplanetary material as proposed by \cite{Sierra2024}, the SPHERE signal may be related to thermal emission from the embedded planet. Further observations are needed to confirm the presence of the SPHERE signal and to determine its origin. We note that due to the close proximity of either of these tentative planet candidates to the inner disk, it is unclear if any one of them could be solely responsible for the opening of the disk gap that we observe. This would depend on their precise mass and orbital configuration, the exploration of which goes beyond the scope of the current study.

Comparison with the planet-hosting disk in the PDS\,70 system may suggest that the presence of spiral arms in 2MASS1612 and the absence of strong grain segregation at the outer gap edge points toward a significantly lower viscosity in the disk around 2MASS1612 compared to PDS\,70. Given the lower mass of the central star in 2MASS1612 (M0.5 vs. G7 spectral type), this may indicate a significantly different parameter regime for planet formation models. A direct detection of the forming planets in 2MASS1612 would thus be of utmost importance as input for planet-formation models.

The inferred mass range and the angular separation of the presumed giant planet in the disk gap make it an excellent target for follow-up observations with the James Webb Space Telescope (JWST). The superior sensitivity of the JWST compared to ground-based observations, combined with the available longer wavelength coverage that allows us to pierce through embedding circumplanetary material would be ideal for directly detecting the thermal signature of the embedded planet. Alternatively, it may be possible to pick up near-infrared or optical accretion signatures of the embedded planet.
The 2MASS1612 system is an excellent example on how near-infrared scattered-light observations can trace planet-disk interaction and guide future surveys for direct planet detections.

\begin{acknowledgements}

This work has made use of data from the European Space Agency (ESA) mission {\it Gaia} (\url{https://www.cosmos.esa.int/gaia}), processed by the {\it Gaia} Data Processing and Analysis Consortium (DPAC,
\url{https://www.cosmos.esa.int/web/gaia/dpac/consortium}). Funding for the DPAC has been provided by national institutions, in particular, the institutions participating in the {\it Gaia} Multilateral Agreement.
SPHERE was designed and built by a consortium made of IPAG (Grenoble, France), MPIA (Heidelberg, Germany), LAM (Marseille, France), LESIA (Paris, France), Laboratoire Lagrange (Nice, France), INAF–Osservatorio di Padova (Italy), Observatoire de Genève (Switzerland), ETH Zurich (Switzerland), NOVA (Netherlands), ONERA (France) and ASTRON (Netherlands) in collaboration with ESO.  SPHERE was funded by ESO, with additional contributions from CNRS (France), MPIA (Germany), INAF (Italy), FINES (Switzerland) and NOVA (Netherlands).  
Additional funding from EC's 6th and 7th Framework Programmes as part of OPTICON was received (grant number RII3-Ct-2004-001566 for FP6 (2004–2008); 226604 for FP7 (2009–2012); 312430 for FP7 (2013–2016)).
P.P. and A.S. acknowledge funding from the UK Research and Innovation (UKRI) under the UK government’s Horizon Europe funding guarantee from ERC (under grant agreement No 101076489).
This paper makes use of the following ALMA data: ADS/JAO.ALMA\#2018.1.00689.S. 
ALMA is a partnership of ESO (representing its member states), NSF (USA) and NINS (Japan), together with NRC (Canada), MOST and ASIAA (Taiwan), and KASI (Republic of Korea), in cooperation with the Republic of Chile. The Joint ALMA Observatory is operated by ESO, AUI/NRAO and NAOJ. 
A.Z. acknowledges support from ANID -- Millennium Science Initiative Program -- Center Code NCN2024\_001 and Fondecyt Regular grant number 1250249.
A.R. has been supported by the UK Science and Technology Facilities Council (STFC) via the consolidated grant ST/W000997/1 and by the European Union’s Horizon 2020 research and innovation programme under the Marie Sklodowska-Curie grant agreement No. 823823 (RISE DUSTBUSTERS project).
R.T. acknowledges funding from the European Research Council (ERC) under the European Union's Horizon Europe research and innovation program (grant agreement No. 101053020, project Dust2Planets).
F.M. acknowledges funding from the European Research Council (ERC) under the European Union's Horizon Europe research and innovation program (grant agreement No. 101053020, project Dust2Planets).
T.B. acknowledges funding from the European Union under the European Union's Horizon Europe Research and Innovation Programme 101124282 (EARLYBIRD) and funding by the Deutsche Forschungsgemeinschaft (DFG, German Research Foundation) under grant 325594231, and Germany's Excellence Strategy - EXC-2094 - 390783311. 
Part of this research was carried out at the Jet Propulsion Laboratory, California Institute of Technology, under a contract with the National Aeronautics and Space Administration. 
This project has received funding from the European Research Council (ERC) under the European Union's Horizon 2020 research and innovation programme (PROTOPLANETS, grant agreement No.~101002188). 
S.F. is funded by the European Union (ERC, UNVEIL, 101076613), and acknowledges financial contribution from PRIN-MUR 2022YP5ACE.
Funded by the European Union (ERC, WANDA, 101039452). Views and opinions expressed are however those of the author(s) only and do not necessarily reflect those of the European Union or the European Research Council Executive Agency. Neither the European Union nor the granting authority can be held responsible for them.
\end{acknowledgements}

\bibliographystyle{aa}
\bibliography{myBib}

\begin{appendix}

\section{Ellipse fitting of the scattered light disk}
\label{app::ellipse-fit}

\begin{figure}
\centering
    \includegraphics[width=0.49\textwidth]{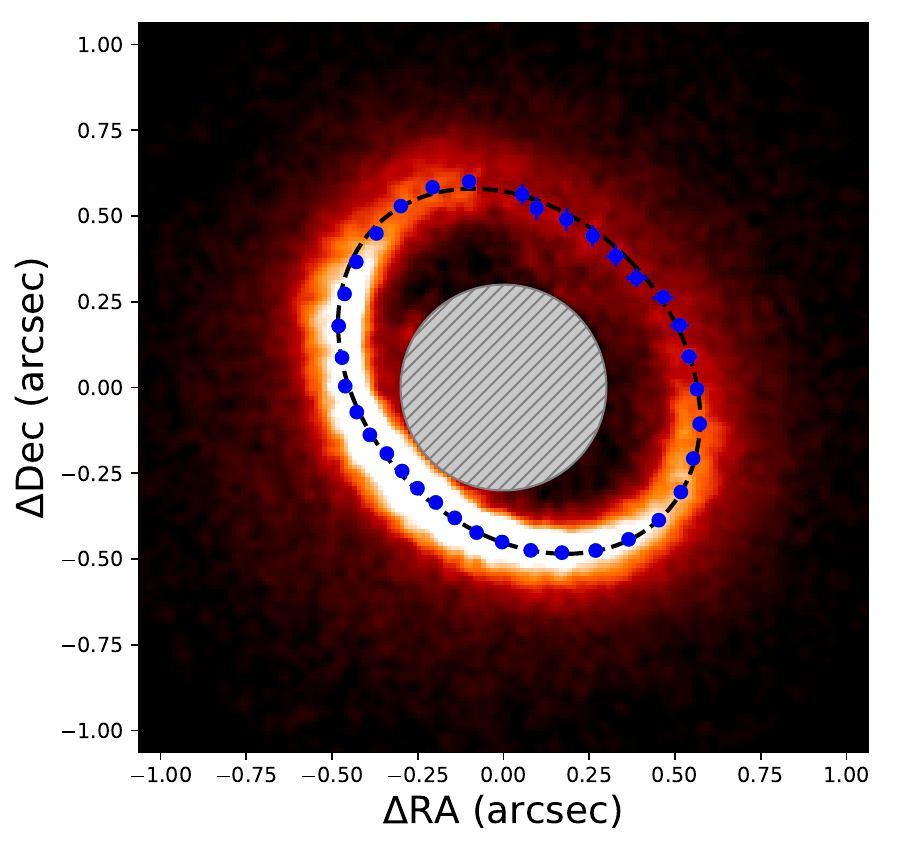}
    \caption{Extracted data points and best fitting ellipse for the outer disk ring. In the cases where they are not clearly visible error bars are roughly similar in size to the data point markers. We indicate the region which was excluded to facilitate a clean fit of the outer ring with a hashed gray circle.}
    \label{fig::ellipse_fit}
\end{figure}

We performed the ellipse fitting on the H-band Q$_\phi$ image as it captures the disk structures with the highest signal-to-noise. 
In order to fit the outer disk ring we first masked the signal of the inner disk to prevent signal confusion for the fitting procedure. The mask we placed has a radius of 24 pixels, i.e. 294\,mas. We then applied the same fitting procedure as described in \cite{Ginski2024} for the ringed systems. This procedure consisted of extracting radial profiles every 1$^\circ$ along the azimuthal direction. For each profile we fitted a Gaussian to find the peak of the outer ring with sub-pixel accuracy. The standard deviation of each Gaussian, normalized by the signal-to-noise ratio of the disk ring at this location, was considered the radial uncertainty of the ring location. Based on the absolute calibration accuracy of the SPHERE frames we used a value of 0.1$^\circ$ as the homogeneous azimuthal uncertainty for each individual data point. This yielded a total of 360 data points for the ellipse fit. Considering the actual image resolution we then binned the resulting data points in bins of 10$^\circ$. The final data points with their uncertainties converted to the Cartesian coordinate system are shown in figure~\ref{fig::ellipse_fit}. We then used these data points as input to run the ellipse fitting algorithm by \cite{oy1998NUMERICALLYSD}. To account for the uncertainty of the data points we use a least-squares Monte Carlo approach (LSMC), i.e. we ran this fitting algorithm 10$^6$ times. For each individual fit we were drawing the location of the individual data points from a normal distribution centered on the original location and with the uncertainty of the data point as standard deviation. The final ellipse parameters are the medians of the resulting distributions and the uncertainties of each parameter are the standard deviation calculated from the individual distributions. We note that the uncertainties in the particular case of 2MASS1612 may be overestimated as we find a median full width at half maximum of the outer ring of 17.2 pixels, i.e. 211\,mas. This is significantly larger than the resolution element in the H-band data of 50\,mas, i.e. the ring itself is radially resolved.

\section{Image de-projection and spiral pitch angles}
\label{app::deprojection}

\begin{figure}
\centering
    \includegraphics[width=0.49\textwidth]{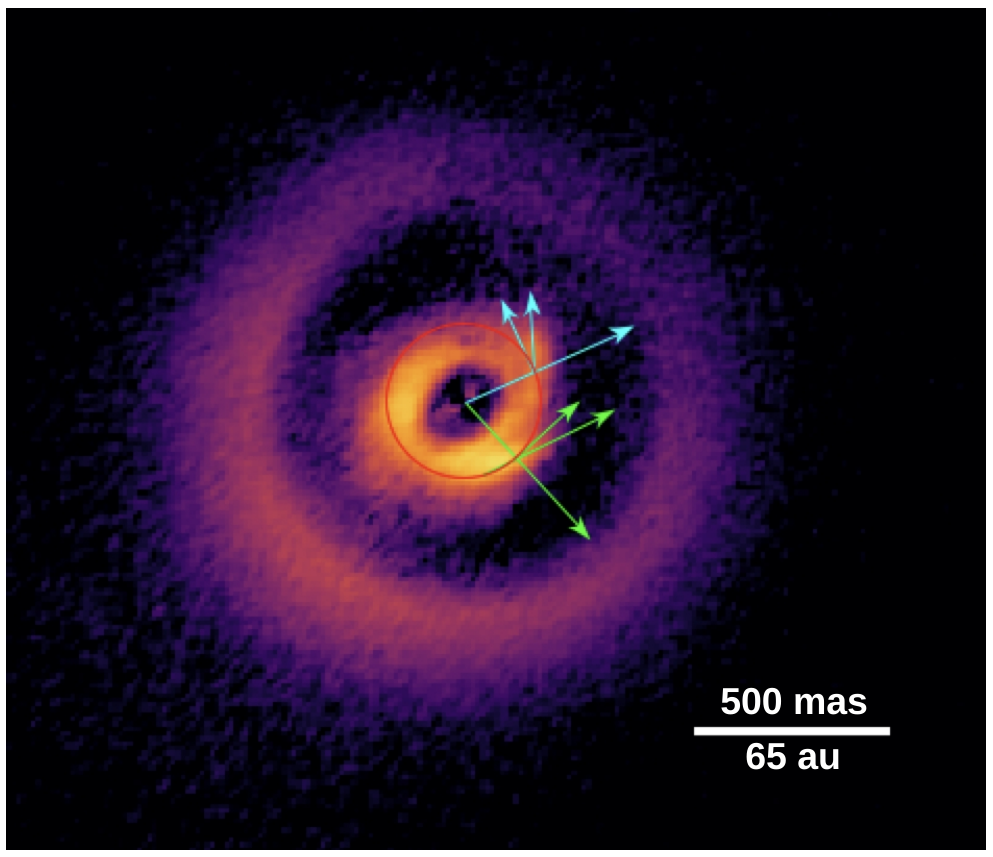}
    \caption{De-projected image of the disk taking into account the measured disk inclination and height and assuming a power-law profile for the disk height structure. The red circle indicates the radial distance at which the spiral pitch angle was measured. For both spirals we indicate with a radial (cyan or green) arrow where we consider the intersection of the red circle with the spiral arm. The azimuthal cyan and green arrows indicate the measured spiral pitch angles for the western and southern arms respectively. This de-projection is based on the H-band data set.}
    \label{fig::deprojected}
\end{figure}

To accurately measure the pitch angles of the spirals in the inner disk we need to de-project the disk to a face on view. This is not trivial as we do not know the precise vertical height profile of the disk. However, the offset of the outer disk ring gives us the aspect ration of the disk the ring location. We then assume that the disk has a flared profile \citep[as seen for other systems by ][]{Ginski2016, deBoer2016, Avenhaus2018}. Since we do not know the precise power-law coefficient for the system, we assume the average power-law found by \citet{Avenhaus2018} for multiple T Tauri stars of 1.2. For the de-projection we first ``flatten'' the disk by correcting for the height-induced offset of the disk from the stellar position along the minor axis. We then stretch the disk along the minor axis to correct for the inclination. The result is shown in figure~\ref{fig::deprojected}. If the de-projection uses a reasonably correct inclination and disk height profile we would expect that the outer ring is circular assuming that the disk is not significantly eccentric. This is indeed the case for the disk outer ring. 
In the same figure we show the vectors used to estimate the pitch angles of the spiral arms. We note that the angle should be somewhat less affected by the disk de-projection for the spiral arm that is along the major axis, compared to the spiral arm along the minor axis.

\section{Hydrodynamic and radiative transfer simulations}
\label{app::hydro}

We performed a series of 3D gas only simulations with the Smoothed Particle Hydrodynamics (SPH) code {\sc phantom} \citep{Price2018}, using $10^6$ particle.  We adopt a stellar mass of $0.75$M$_\odot$ and set the accretion radius to 1\,au. We setup a disc with an initial gas mass of 0.005\,M$_\odot$ between 5 and 210\,au, a tapered surface density profile:
$$
\Sigma(r) = \Sigma_c \left( \frac{r}{r_c}\right)^{-\gamma} \exp\left( -\left( \frac{r}{r_c}\right)^{2-\gamma} \right)
$$
with $\gamma=-1$ and $r_c=150$\,au. We adopted a vertically isothermal equation of state with $H/R=0.1$ at $r=50$\,au and sound speed power-law index of -0.25. We set the shock viscosity $\alpha_{\rm av}=0.2$ to obtain a mean Shakura-Sunyaev viscosity of $5\times 10^{-3}$.

We embedded a planet at 50\,au with an initial mass of either 1,2,5 or 10\,M$_\mathrm{Jup}$, with accretion radius set to 0.125 times the Hill radius. We evolved the simulations for 45 planet orbits, by which time the planets reached a mass of 1.4, 2.6, 5.9, 11.1 \,M$_\mathrm{Jup}$, and migrated to radii between 47 and 49\,au.

We post-processed the models with the radiative transfer code {\sc mcfost} \citep{Pinte2006,Pinte2009} to compute the dust temperature structure and synthetic images.

We used a distance of 132\,pc \citep{GaiaDR3} and 5\,Myr isochrones to set the star \citep{Siess2000} and planet \citep{Allard2012} luminosities and effective temperature, giving a radius of 1.55\,R$_\odot$, and $T_{\rm eff} = 4000$\,K for the central star. For the planets, we obtained radii of 0.15, 0.16, 0.17, and 0.21\,R$_\odot$ and $T_{\rm eff} = 850$, 11250, 1670, and 2130\,K.
We assumed astrosilicate grains \citep{Weingartner2001} with sizes following
$\mathrm{d}n(a) \propto a^{-3.5}\mathrm{d}a$ between 0.03 to
1000$\mu$m, a gas-to-dust ratio of 100, and computed the dust optical properties using Mie theory. 

\end{appendix}
\end{document}